\documentstyle[11pt,a4wide,epsf]{article}
\newcommand{\be}{\begin{equation}}
\newcommand{\ee}{\end{equation}}
\newcommand{\etal}{{\it et al.}}
\def\vec#1{{\left| {#1} \right\rangle}}
\def\D{\frac{d}{dz}}

\makeatletter

\@addtoreset{equation}{section}
\def\section{\@startsection {section}{1}{\z@}{-3.5ex plus -1ex minus
 -.2ex}{2.3ex plus .2ex}{\large\bf\centering}}
\def\subsection{\@startsection{subsection}{2}{\z@}{-3.25ex plus%
 -1ex minus -.2ex}{1.5ex plus .2ex}{\bf}}
\def\subsubsection{\@startsection{subsubsection}{3}{\z@}{-3.25ex plus%
 -1ex minus -.2ex}{1.5ex plus .2ex}{\sl}}
\makeatother

\begin{document}
\def\ss{\scriptstyle}
\baselineskip 17pt
\parindent 12pt
\parskip 9pt

{
\parskip 0pt
\newpage
\begin{titlepage}
\begin{flushright}
DAMTP--96--19\\
KCL--TH--96--6\\
22 March 1996\\[5cm]
\end{flushright}
\begin{center}
{\Large{\bf
A crossing probability for critical percolation in two dimensions
}}\\[1.5cm]
{\large G.M.T.~Watts
}\footnote{%
e-mail: gmtw@mth.kcl.ac.uk }\\[8mm]
{\em Department of Mathematics,}\\
{\em King's College London,}\\
{\em Strand, London, WC2R 2LS, U.K.}\\[5mm]
{\bf{ABSTRACT}}
\end{center}
\begin{quote}

Langlands et al.\ considered two crossing
probabilities, $\pi_h$ and $\pi_{hv}$, in their extensive numerical
investigations of critical percolation in two dimensions.
Cardy was able to find the exact form of $\pi_h$ by treating it as a
correlation function of boundary operators in the $Q\to 1$ limit of the
$Q$ state Potts model. We extend his results to find an analogous
formula for $\pi_{hv}$ which compares very well with the numerical
results.

\end{quote}
\vfill
\end{titlepage}
}

\section{Introduction}
\label{sec:one}
\setcounter{footnote}{0}

Critical percolation in two dimensions has been intensively studied
numerically by Langlands \etal, \cite{Yvan1,Yvan2}.
One of the results they find is that various crossing
probabilities, $\pi_h$, $\pi_{hv}$ (to be defined below) are invariant
under conformal transformations of the plane.
In \cite{Cardy}, by combining a standard identification of critical
percolation and the 1 state Potts model with boundary conformal field
theory techniques, Cardy was able to find the exact form of $\pi_h$.
There are three steps needed to extend this to $\pi_{hv}$: 1. Show
that $\pi_{hv}$ corresponds to some particular boundary conditions in
the 1 state Potts model, and thus is given by a correlation function
of conformal field theory boundary operators; 2. identify theses
boundary operators; 3. deduce a differential equation for the
correlation function, and compare the result with $\pi_{hv}$.
In this letter we complete steps 1 and 3. Of course this is far from
being a complete derivation of $\pi_{hv}$, but having shown
that it is given by a correlation function of boundary operators, then
one knows that $\pi_{hv}$ must satisfy one of a discrete set of
known differential equations, and examining the simplest of these we
find a single function which has suitable asymptotic behaviour.
Comparing this function with the numerical data for $\pi_{hv}$
presented in \cite{Yvan2}, our agreement is as good as that of Cardy's
formula for $\pi_h$.

\section{Critical percolation and boundary conformal field theory}
\label{sec:two}

For an extremely good presentation of critical percolation, and the
full content of the hypotheses of universality and conformal
invariance, we refer the reader to \cite{Yvan1}. At its simplest, one
can envisage a rectangle, and take as observables all combinations of
probabilities of crossing from one set of points on the boundary to
another disjoint set. The hypothesis of conformal invariance is that
such probabilities are invariant under conformal transformations of
the plane. For example, the probabilities, $\pi_h$, of crossing from
the left side to the right, and $\pi_{hv}$ of being able to cross
simultaneously from left to right and from top to bottom, depend only
on the aspect ratio $r$ of the rectangle. It is easy to see that, with
$r = \hbox{height}/\hbox{width}$,
\be
  \pi_{hv}(r) = \pi_{hv}(1/r)
\;,\;\;\;\;
  \pi_{hv}(r) / \pi_h(r) \to 1 \hbox{ as } r \to 0
\;.
\label{eq:bcs}
\ee
Langlands et al.\ provide numerical values for $\pi_h$ and $\pi_{hv}$
for aspect ratios $0.136 < r < 7.351$.

The property of the crossing probabilities described above, that they
are invariant under conformal mappings of the plane, is that of
correlation functions of $h=0$ boundary operators in a $c=0$ conformal
field theory.  (For a review of boundary operators in conformal field
theory, see e.g. \cite{Cardybc,Cardyopc}). It was in this way
that Cardy was able to find a formula for $\pi_h(r)$.
It will be instructive to repeat the relevant points of his derivation
here.

The combination $(1 - \pi_h)$  may be identified with the
$Q \to 1$ limit of the $Q$ state Potts model partition function with
specific boundary  conditions: namely free boundary conditions on the
vertical sides of the rectangle, and the spins fixed to different
values on the horizontal sides. One can identify this partition
function as the correlation function of four operators, placed at the
corners of the rectangle, which change the boundary conditions.

By considering the 2 and 3 state Potts models, for which the
correspondence with $c=1/2$ and $c=4/5$ conformal field theories is
very well known, Cardy could identify the relevant boundary changing
operators as type $(1,2)$ boundary primary fields. This implied that the
correlation function of four such operators should satisfy a second
order differential equation. This equation is very easy to solve, and
by taking the appropriate combination of the two independent solutions
of this equation, Cardy could fit the data of \cite{Yvan1} to a very
high degree of accuracy. In fact, this fit was so good, that
Langlands \etal\ decided to adopt Cardy's result as $\pi_h$, rather
than their own numerical data, in their later paper on more
complicated calculations \cite{Yvan2}.

\section{Boundary conditions for $\pi_{hv}$}

One can similarly find $\pi_{hv}$ from the $Q \to 1$ limit of the $Q$
state Potts model partition function with certain other boundary
conditions which, exactly as for $\pi_h$, are not themselves defined
for $Q=1$. There is a standard argument relating the partition
function in the Potts model to a sum over a graphical expansion
\cite{FK}. In this, one considers all sets of `clusters' $\xi$, where
a cluster is a connected set of bonds on a lattice, and no two clusters
intersect. Then the partition function may be written as a sum over
clusters, with weights which depend on the type of cluster,
corresponding to the type of boundary conditions one wishes to
impose. For our purposes, it will be enough to consider a partition
function defined for a square lattice on a rectangle with the sides of
the rectangle consecutively numbered 1 to 4, and to consider an
expansion of the form
\be
  Z(a,b,c,d,Q) = \sum_\xi
      p^{B(\xi)}
      (1-p)^{B - B(\xi)}
      a^{N_a(\xi)}
      b^{N_b(\xi)}
      c^{N_c(\xi)}
      d^{N - N_a(\xi) - N_b(\xi) - N_c(\xi)}
\;,
\label{eq:z}
\ee
where $p$ is the normalised Boltzmann weight for an edge connecting
sites with the same spin, and which will be set to the critical
probability $p_c=1/2$ for a bond to be open, when we finally take 
$Q \to 1$. In (\ref{eq:z}), $B$ is the total number of bonds, $B(\xi)$
is the number of bonds in $\xi$, $N$ is the total number of clusters,
$N_a, N_b$ and  $N_c$ are the numbers of clusters intersecting no
sides of the rectangle, exactly 1 side and two adjacent sides
respectively, and $a$, $b$, $c$ and $d$ are weights which depend on
the boundary conditions. Taking $a=b=c=d=Q$ one recovers the $Q$ state
Potts model partition function for free boundary conditions, and
$Z(1,1,1,1,1) = 1$ in the 1  state Potts model. However, if one can
arrange boundary conditions such that, as $Q \to 1$, that $a \to 1$, 
$b \to 1$ and  $c \to 1$ and $d \to 0$, then one will recover the
probability that no cluster intersects non-adjacent boundaries, 
i.e.\ $1 -\pi_h -\pi_v +\pi_{hv}$. 

Let us consider the following boundary condition, which we denote by
$(AB\ldots C)$: in this boundary condition the spin at each site in
the boundary may take its value freely in any of the states
$A,B,\ldots C$. If we assign the  conditions $b_i$ to the four sides
of the rectangle in turn,
\[
 b_1 = (A,B,X_{11},X_{12}, \ldots,X_{1n} )
\;,\;\;
 b_2 = (B,C,X_{21},X_{22}, \ldots,X_{2n} )
\;,\;\;
\]
\be
 b_3 = (C,D,X_{31},X_{32}, \ldots,X_{3n} )
\;,\;\;
 b_4 = (D,A,X_{41},X_{42}, \ldots,X_{4n} )
\;,\;\;
\ee
where all the spins $\{X_{ij}\}$ and $\{A,B,C,D\}$ are distinct, then
we see that the partition function of the $Q$--state Potts model with
these boundary conditions will be in the form (\ref{eq:z}) with
weights $a = Q$, $b = n+2$, $c=1$ and $d=0$. Then, if we take $n =
(Q-5)/4$ (which is perfectly valid for $Q = 5,9,\ldots$) and then $Q
\to 1$, we  will recover the required weights for our crossing
probability, i.e.
\be
 \lim_{Q \to 1}
  Z(Q, (Q+3)/4,1,0,Q)
= 1 - \pi_h - \pi_v + \pi_{hv}
= \pi_{hv}
\;,
\label{eq:zlim}
\ee
where we take for granted the universality of critical percolation and
set $\pi_h + \pi_v = 1$ as is the case for percolation by sites on a
triangular lattice. 

Unfortunately, the boundary changing operator which changes the
boundary conditions from type $(A,B,X_{11},X_{12}, \ldots,X_{1n} )$ to
$(B,C,X_{21},X_{22}, \ldots,X_{2n} )$ does not exist in the
2 state Potts model for any $n$, and has only been identified to date
for $n=0$ in the 3 state Potts model \cite{Cardybc,salb}, and
consequently  we can not deduce a differential equation for $Z$.
However, on the basis of this argument, we can certainly expect that
$\pi_{hv}$ is expressible as some correlation function of four $h=0$
conformal boundary primary fields.

\section{Differential equations for correlation functions}

Correlations of conformal primary fields $\phi_h(z)$ can satisfy
differential equations, one differential equation for each null state
in the highest weight representations of the Virasoro algebra with
highest weight state
$ \phi_h(0) \vec 0$, and the general method by which such differential
equations are derived is given in \cite{BPZ}.
Since the space of null vectors is a highest weight space, one can
restrict attention to the differential equations arising from the
vanishing of highest weight states. For $h=c=0$ it is easy to compute
the highest weight states with low conformal weight in the Verma
module and we give the first four of these in table \ref{tab:nulls}.
{
\renewcommand{\arraystretch}{1.6}
\headsep -0.5in
\begin{table}[e]
\caption{Low lying highest weight vectors in the $h=c=0$ Verma module.}
\label{tab:nulls}

\[
\begin{array}{|c|@{~~}l@{~~}|}
\hline
  \hbox{level }n     & \hbox{Null vector } {\cal N}(n) \\
\hline\hline

  1 & L_{-1} \vec 0 \\
  2 & \left(\, L_{-1}  L_{-1} - (2/3) L_{-2} \, \right) \vec 0 \\

  5 & \left(\, L_{-1}^3 - 6 L_{-2}L_{-1} + 6 L_{-3} \,\right) {\cal N}(2) \\
  7 & \left(\,
  \begin{array}{c}
  L_{-1}^5 - (40/3) L_{-2}L_{-1}^3 + (256/9) L_{-2}L_{-2}L_{-1}
  + (52/3) L_{-3}L_{-1}L_{-1} \\
  - (256/9) L_{-3}L_{-2} - (104/3) L_{-4}L_{-1}  + (208/9) L_{-5}
  \end{array}
  \,\right) {\cal N}(2) \\[6mm]
\hline\end{array}
\]
\end{table}
}
The four-point functions of four $h=0$ boundary primary fields will
only depend on the cross ratio $\eta$, of their coordinates, so we
will write
\[
  \langle\,\phi(z_1)\, \phi(z_2)\, \phi(z_3)\, \phi(z_4)\, \rangle
= F(\eta)
\]
where $\eta = ((z_3 - z_4)(z_2 - z_1)/((z_3 - z_1)(z_2 - z_4))$.
The differential equations arising from the first three null vectors
are equally easy to derive, and are given, along with their solutions,
in table \ref{tab:des}.%
{\begin{table}[e]
\renewcommand{\arraystretch}{1.8}
\headsep -0.5in
\caption{Differential operators and solutions}
\label{tab:des}

\[
\begin{array}{|c|@{~~~~}l@{~~~~}|}
\hline
  \hbox{level }n     &
   \\
\hline\hline

  1 & \D \\
  F^{(1)}
    & c_1 \\ \hline

  2 & (z(z-1))^{-2/3} \;\;\D\;\; (z(z-1))^{2/3} \;\;\D  \\
  F^{(2)}
    & {\displaystyle
        c_1
  \;+\; c_2 \int_0^z \frac{dx}{( x(1-x) )^{2/3} }
      }\\[4mm] \hline

  5 & (z(z-1))^{-2} \;\; \frac{d^3}{dz^3}\;\; (z(z-1))^{4/3}
                   \;\;\D\;\; (z(z-1))^{2/3} \;\;\D   \\
  F^{(5)}
    & {\displaystyle
        c_1
  \;+\; \int_0^z \frac{dx}{( x(1-x) )^{2/3} }
      \left[ c_2 \;+\;
            \int_0^x dt \frac{ c_3 + c_4 t + c_5 t^2 }{(t(1-t))^{4/3}}
      \right]      } \\[4mm]

\hline\end{array}
\]
\end{table}
}
The conditions (\ref{eq:bcs}) translate into
\be
  F(\eta) = F(1 - \eta)
\;,\;\;\;\;\;\;
     {{F(\eta)}\over{ \eta^{1/3}\; {}_2 F_1(1/3,2/3;4/3;\eta)}}
\to \frac{3 \Gamma(2/3)}{\Gamma(1/3)^2}
\;
\hbox{ as $\eta \to 0$}
\;.
\label{eq:bcs2}
\ee
It is impossible to find a solution of type $F^{(1)}$ or $F^{(2)}$
which satisfies (\ref{eq:bcs2}), but there is a unique
solution of type $F^{(5)}$ which does, and which can be presented
variously as
\begin{eqnarray}
    \!\!\!\!\!\!\!\!\!\!{\cal F}(z)
&=& \frac{ \Gamma(2/3)}{\Gamma(1/3)^2}
    \int_0^z \frac{dt}{[ t(1-t)]^{2/3}}
\;\;-\;\;
    \frac 2{3 \Gamma(1/3)\Gamma(2/3)}
    \int_0^z \frac{dt}{[t(1-t)]^{2/3}}
    \int_0^t \frac{du}{[u(1-u)]^{1/3}}
\nonumber\\
&=& \frac{ \Gamma(2/3)}{\Gamma(1/3)^2}
    \int_0^z \frac{dt}{[ t(1-t)]^{2/3}}
\;\;-\;\;
    \frac 1{ \Gamma(1/3)\Gamma(2/3)}
    \int_0^z \frac{dt}{[t(1-t)]^{2/3}}
    \,{}_2F_1(1,4/3;5/3;t)
\nonumber\\
&=& \frac{3 \Gamma(2/3)}{\Gamma(1/3)^2}
    z^{1/3}\,{}_2F_1(1/3,2/3;4/3;z)
\;\;-\;\;
    \frac z{ \Gamma(1/3)\Gamma(2/3) }
    \;\;{}_3F_2(1,1,4/3;2,5/3;z),
\label{eq:f5}
\end{eqnarray}
where ${}_2F_1$ is the standard hypergeometric function, and ${}_3F_2$ is a
generalised hypergeometric function.

\section{Comparison with data and discussion}

The case treated in \cite{Yvan2} is of the crossing
probabilities defined for a rectangle of aspect ratio $r$.
If we take the points $z_j$ to be at $(-k^{-1},-1,1,k^{-1})$, then we
can map these to the corners of a rectangle by
\[
   w = \int_0^z \frac{ dx}{\sqrt{(1 - x^2)(1 - k^2x^2)}}
\;,
\]
in which case the aspect ratio of the rectangle is given by
$r=K(1-k^2)/2K(k^2)$, where $K(u)$ is the complete elliptic integral
of the first kind.
The cross ratio of these four points is
$\eta = ((1-k)/(1+k))^2$.

In figure 1 we plot $ \log(\pi_{hv}) $ against $ \log(r) $
for both the numerical results obtained in \cite{Yvan2} and for the
function ${\cal F}(x)$ above.
The agreement is excellent%
\footnote{In fact Langlands et al.\ only computed $\pi_{hv}$ for
$r<1$, since it is invariant under $r \to 1/r$, and consequently
half of the data points we give are duplicated.}.

There are still several points which remain unclear in the results
presented here.
Firstly, the integral expression of $\pi_h$ found by Cardy can be
easily identified as one of the standard solutions of Dotsenko and
Fateev \cite{DF}. However, despite the fact that ${\cal F}$ satisfies
the standard differential equations for a type $(3,4)$ primary field,
we have not yet been able to find a way to express (\ref{eq:f5}) in
Dotsenko--Fateev form. 
If one allows more general vertex operator constructions, such as
for $W$-algebras, then it is easy to see how the troublesome second
term in ${\cal F}$ might arise as a screened four-point function: one
way, using purely bosonic vertex operators, is
\[
\lim_{w \to \infty} \;\;
\int_{0}^z\! dt \, 
\int_{0}^t\! du \;\;
\langle 3\alpha+\beta+\gamma |
\; V_\alpha(w) 
\; V_\alpha(1) 
\; V_0(z)
\; V_\beta(t)
\; V_\gamma(u)
\; V_\alpha(0)
\; | 0 \rangle
\;,
\]
where $\alpha^2 = 1/2$, $\alpha\!\cdot\!\beta = -2/3$,
$\alpha\!\cdot\!\gamma = -1/3$, 
$\beta\!\cdot\!\gamma=0$ and where 
$V_\lambda(z) \equiv :\exp(i \lambda\cdot\phi(z)):$ is a bosonic 
vertex operator, $V_\alpha$  represents a weight 0 field, and $\int
V_\beta$ and $\int V_\gamma$ are screening charges. However, we have
not yet found any pressing case for such an interpretation for any
particular extended algebra.

Secondly, although ${\cal F}(z)$ satisfies the fifth order
differential equation in table \ref{tab:des}, it also satisfies the
third order differential equation
\be
  \D \;\; (z(z-1))^{1/3} \;\;\D\;\; (z(z-1))^{2/3} \;\;\D \;\; {\cal F}
= 0
\;.
\label{eq:3de}
\ee
This is a much more appealing equation as the three independent
crossing probabilities $1, \pi_h, \pi_{hv}$ span the solutions to
(\ref{eq:3de}), but this equation does not arise from the vanishing of
a vector at level 3 in the $h=0$ Virasoro Verma module.
It would be very nice indeed if this equation could be
derived from a null--vector vanishing for some generalisation of the
Virasoro algebra, such as those in \cite{Saleur,Honec}, but we have
not been able to do this yet.

\section{Conclusions}

{}From a simple extension of Cardy's method of \cite{Cardy}, we have
found an excellent candidate for the crossing probability $\pi_{hv}$
which was investigated numerically by Langlands et al.\ in
\cite{Yvan2}.
Unfortunately, we have not been able to find a real derivation of our
result, but work is in progress which should elucidate rather more
the nature of the conformal field theory underlying this result.

\newpage
{\bf Acknowledgments}

\noindent
I would very much like to thank Yvan Saint-Aubin for many discussions
on percolation and crossing probabilities, and lattice models,
conformal field theory and partition functions in general, and
J.~Cardy and G.~Grimmett for very interesting discussions on this
problem,  and for their comments on earlier drafts.

I would also like to thank H.~Kausch for many useful discussions on the
structure of $c=0$ conformal field theory, and the many bizarre
possibilities which arise in this case, and W.~Eholzer for drawing
reference \cite{Honec} to my attention.

GMTW is supported by an EPSRC advanced fellowship.

\epsfxsize=\hsize
\epsfbox{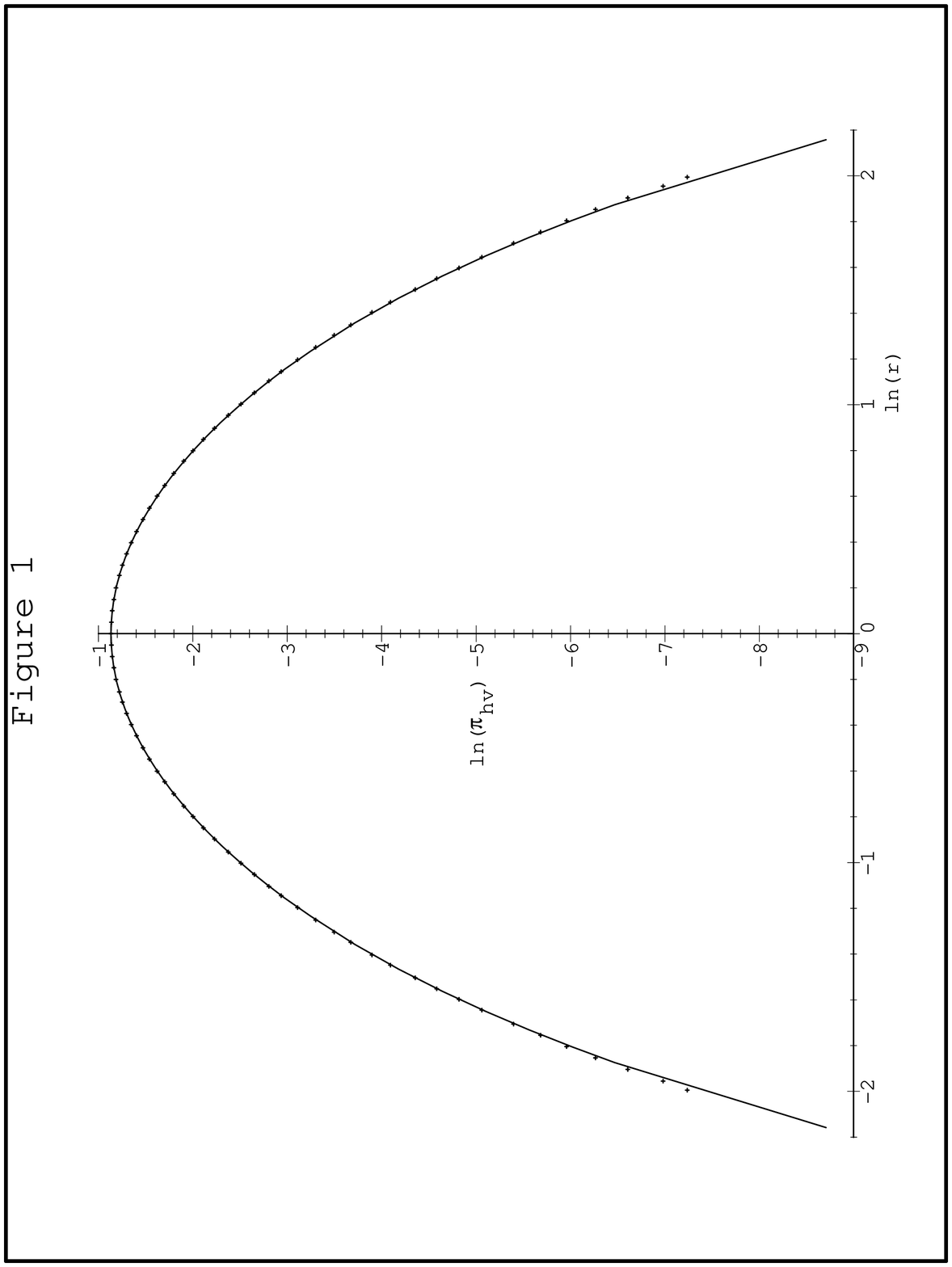}

\end{document}